\documentclass[11pt]{article}

\usepackage{amsmath,amsfonts,amsthm,epsf,stackrel}

\usepackage{graphicx,subfigure}
\usepackage{natbib}
\usepackage{setspace}
\makeatletter
\usepackage{enumerate}
\usepackage{float}

\RequirePackage{natbib}

\usepackage[margin=2cm]{geometry}

\usepackage{tcolorbox}

\usepackage{dsfont}
\usepackage{multirow}

\usepackage{booktabs} 

\usepackage{hyperref}
\hypersetup{
	colorlinks=true,
	linkcolor=blue,
	filecolor=magenta,      
	urlcolor=cyan,
	pdftitle={Overleaf Example},
	pdfpagemode=FullScreen,
}

\newcommand{\ignore}[1]{}

\newcommand{\tbl}[1]{\textcolor{black}{#1}}

\newtheorem{definition}{Definition}[section]

\newtheorem{OP}{Optimization Problem}

\begin{document}
	
	\title{Optimal confidence interval for the difference between proportions}
	\author{
		Almog Peer, David Azriel\\
		Technion - Israel Institute of Technology
	}
	\date{}
	\maketitle
	
		
		%
		%
		%
	
	\begin{abstract}
		Estimating the probability of the binomial distribution is a basic problem, which appears in almost all introductory statistics courses and is performed frequently in various studies. In some cases, the parameter of interest is a difference between two probabilities, and the current work studies the construction of confidence intervals for this parameter when the sample size is small. Our goal is to find the shortest confidence intervals under the constraint of coverage probability \textcolor{black}{being at least as large as a predetermined level.} For the two-sample case, there is no known algorithm that achieves this goal, but different heuristics procedures have been suggested, and the present work aims at finding optimal confidence intervals.
		In the one-sample case, there is a known algorithm that finds optimal confidence intervals presented by Blyth and Still (1983). It is based on solving small and local optimization problems and then using an inversion step to find the global optimum solution. 
		We show that this approach fails in the two-sample case and therefore, in order to find optimal confidence intervals, one needs to solve a global optimization problem, rather than small and local ones, which is computationally much harder. We present and discuss the suitable global optimization problem. Using the Gurobi package we find near-optimal solutions when the sample sizes are smaller than 15, and we compare these solutions to  some existing methods, both approximate and exact. We find that the improvement in terms of lengths with respect to the best competitor varies between 1.5\% and 5\% for different parameters of the problem. Therefore, we recommend the use of the new confidence intervals when both sample sizes are smaller than 15. Tables of the confidence intervals are given in the Excel file in this \href{https://technionmail-my.sharepoint.com/:f:/g/personal/ap_campus_technion_ac_il/El-213Kms51BhQxR8MmQJCYBDfIsvtrK9mQIey1sZnZWIQ?e=hxGunl}{link}.
	\end{abstract} 
	
	
	\section{Introduction}
	The task of constructing confidence intervals for the proportion of 
	the binomial distribution is a basic problem in statistics, which appears in almost 
	all introductory statistics courses and is performed frequently in many studies.
	In some cases, the parameter of interest is the difference between two proportions, and the present work studies the construction of confidence intervals for this parameter. Specifically, if $p_1$ and $p_2$ are two proportions, the parameter of interest is $\Delta= p_1-p_2$. Other functions, such as the ratio $p_1/p_2$ or the log odds ratio $\log\left(\frac{{p_1}/(1-p_1)}{p_2/(1-p_2)}\right)$ will not be discussed here, but we believe that our methodology can be extended to these functions. \tbl{For a discussion about comparisons among different functions of $p_1,p_2$ see \citet{Brumback2008}.}
	\textcolor{black}{Our aim in this work is to study confidence intervals with minimal length. Even when the sample size is small, this is not a trivial problem, and we show below how computational difficulties can be overcome and present (almost) optimal confidence intervals for the stated problem. }

	First, one needs to distinguish between an exact confidence interval (henceforth, CI) and an approximate CI. An exact CI has a guarantee that the \textcolor{black}{coverage probability} is above some predetermined level of $1-\alpha$ for all the parameter space, while \textcolor{black}{an} approximate CI  achieves this level only asymptotically, and might have a smaller coverage probability for some values of the parameter. An exact CI has the advantage of guaranteeing the desired level for every sample size and for every value of the parameter. However, it might come at the cost of \textcolor{black}{wider intervals}. 
	This work focuses on exact CI and small sample sizes.

	We now review some widely-used methods for the one-sample case. The most popular one is the Wald CI, which is based on the normal approximation of the binomial variable. Specifically, let $X\sim Binomial(n,p_1)$, and let $\hat{p}_1=X/n$. Wald CI is $\hat{p}_1 \pm  z_{1-\frac{\alpha}{2}} \sqrt{\frac{\hat{p}_1(1-\hat{p}_1)}{n}} $,
	where $z_{1-\frac{\alpha}{2}}$ is the $1-\frac{\alpha}{2}$ quantile of the standard normal distribution. 
	The Wald CI is symmetric around the observed proportion $\hat{p}_1=X/n$, and its width depends on the variance estimator and the level of confidence  $1-\alpha$. 
	Among the approximate CIs, the Wilson score (\citet{wilson1927probable}) gained some popularity. Similar to the Wald CI, the Wilson CI is based on the normal approximation, but with a different variance estimator. 
	\citet{agresti1998approximate} showed that the performance of the Wald CI is much inferior to the Wilson CI in terms of coverage probability . 
	Agresti and Coull also suggest another CI, which they call an adjusted Wald CI. The idea is to simply take $X^*=X+2,n^*=n+4$ and compute the Wald CI with $X^*$ and $n^*$. 
	
	\citet{brown2001interval} provided a comprehensive review of different methods to construct CIs. They compared performance in terms of 
	minimum coverage level, average coverage level, and average diversion from $1-\alpha$.
	Based on the above criteria, they recommended the Wilson score CI or the Jeffreys CI for $n < 40$. The Jeffreys CI is obtained by using a prior $BETA(\frac{1}{2},\frac{1}{2})$, known as the Jeffreys prior, and taking the middle $1-\alpha$ area under the posterior distribution.
	For $n \geq 40$, Brown et al. suggested using  either the Wilson or the Jeffreys CIs or the Agresti Coull method that was mentioned above. 
	
	The first exact CI for the one-sample case was suggested by \citet{clopper1934use}, and it is the intersection
	of two one-sided CIs. The Clopper and Pearson CI is generally too conservative - the intervals are fairly wide.
	Correspondingly, the coverage probability  is higher than the desired level, especially for small $n$. 
	\citet{sterne1954some} developed an exact CI that is shorter than the Clopper and Pearson CI, and is optimal in the sense of \textcolor{black}{having a minimal length, i.e., the sum of $n+1$ confidence regions is minimal among all CIs with} the correct coverage probability. However, 
	\citet{crow1956confidence} showed that the Sterne method might lead to confidence regions that are the union of intervals and not a single interval. 
	Crow further modified the Sterne method to return only confidence regions consisting of one interval for any $x$, preserving the above optimality property for CIs.
	\citet{blyth1983binomial} proposed an algorithm
	that finds all optimal CIs that are intervals, including 
	the Crow CI. In Section \ref{sec:BS} the Blyth and Still algorithm is described in detail, as we wish to generalize it to the two-sample case. \textcolor{black}{\cite{blaker2000confidence} show that the Blyth and Still CI does not maintain nestedness. This means that for two given coefficients $1-\alpha >1-\alpha'$, the corresponding CI of coefficient $1-\alpha'$ is not necessarily contained in the other CI. Therefore, Blaker proposed an algorithm for constructing a CI that is always an interval and maintains nestedness but is not necessarily optimal.}

Now, we will review several CIs for the two-sample case, i.e., for $\Delta=p_1-p_2$. The 
Wald CI can be easily generalized based on the normal 
approximation of the differences of the averages.
Specifically, let $X\sim Binomial(n,p_{1}),Y\sim Binomial(m,p_{2})${, where $X$ and $Y$ are independent and }
let $\hat{p}_{1}=X/n, \hat{p}_{2}=Y/m$. The Wald CI for $\Delta$ is
\[
\hat{p}_{1}-\hat{p}_{2} \pm  z_{1-\frac{\alpha}{2}} \sqrt{\frac{\hat{p}_{1}(1-\hat{p}_{1})}{n}+
	\frac{\hat{p}_{2}(1-\hat{p}_{2})}{m}.
}
\]
\citet{miettinen1985comparative}
demonstrated the poor coverage of this CI in a few examples and suggested 
relying on more stabilized estimators of the variance, which are based on quantiles of the chi-square distribution, and result in an approximate CI. \textcolor{black}{Recently, \cite{Antonio2024} revisited the work of \citet{miettinen1985comparative} and suggested a bias correction factor in the context of hypothesis testing. }

\citet{newcombe1998interval} reviewed 11 methods for creating CIs for $\Delta$, including the methods that were mentioned above.
Newcombe compared the methods by the average coverage, the minimal coverage, and the percentage of non-coverage.  
{Furthermore,} Newcombe suggested a method{of his own called} `hybrid score' and it performed well in the above criteria; see Section \ref{sec:list} for more details.
Another recommended method for constructing a CI for $\Delta$ was proposed by \citet{agresti2000simple}. Generalizing Agresti and Coull CI, they proposed to add four pseudo observations, one to each group, i.e., 
define $X^*=X+1, Y^*=Y+1,n^*=n+2, m^*=m+2$
and then calculate the Wald CI for the difference.

A few exact CIs for $\Delta$ were also developed. 
\citet{santner1980small} proposed 
three different methods to construct exact CIs. One of them, called the tail method, has gained popularity due to its simplicity and ease of calculation. This method can be thought of as a two-dimensional analog of the Clopper and Pearson CI for one proportion, where the CIs are an intersection of two one-sided intervals. This method typically leads to too conservative intervals, as shown by \citet{chan1999test}. The latter paper suggests a different method for constructing exact CIs.
\citet{agresti2001small} studied exact CIs for the two-sample case. 
They reviewed the Chan and Zhang CI and suggested a modification that results in significant improvement in performance.
The method is described in detail in Section \ref{sec:list}.
\citet{fagerland2015recommended} compared several methods including the ones mentioned above and also others both approximate and exact. Their main criterion for comparison was the closeness to the nominal level $1- \alpha$.
They recommended using the Agresti and Min CI. {\citet{Fay2021} reviewed different methods and compared them both in terms of testing and confidence intervals. They found that no one method can meet all the desirable properties and provide recommendations based on which properties are given more importance. A related topic is exact tests in $2\times2$ contingency tables; see \citet{Keer2023} and references therein. This work uses optimization methods to maximize
	the number of outcomes in a rejection region similarily to what is done in the current paper.}

To sum up, for the one-sample case there exists an algorithm that minimizes the \textcolor{black}{sum of interval's} length of the CI under the constraint of obtaining a certain coverage level. For the two-sample case, such an algorithm does not exist, but rather different heuristics were suggested. This work aims at filling this gap, namely,{to construct} an algorithm that computes the optimal CI for small sample sizes in the two-sample case and to compare it to existing methods.   

The rest of the work is organized as follows: in Section \ref{probleam statement} the optimization problem is stated and the basic notation is introduced.  The algorithm  suggested in \citet{blyth1983binomial} finds the optimal solutions for the one-sample case. It is based on solving small and local optimization problems and then using an inversion step to find the global optimum solution. Section \ref{BSG} presents the algorithm and discusses extensions to the two-sample case. It is shown that this approach fails in the two-sample case and therefore, in order to find{an} optimal CI, one needs to solve a global optimization problem, rather than small and local ones, which is computationally much harder. The global optimization problem is presented and discussed in Section \ref{sec:full}. Using the \citet{gurobi} package, we find near-optimal solutions when the sample sizes are smaller than 15, and we compare these solutions to  some existing methods, both approximate and exact in Section \ref{sec:comparsions}. We find that the improvement in terms of lengths with respect to the best competitor varies between 1.5\% and 5\% for different parameters of the problem. Section \ref{sec:discussion} concludes with some recommendations and future research directions.

\section{Problem statement}\label{probleam statement}

Recall that \textcolor{black}{$X$ and $Y$ are independent,} $X\sim Binomial \{(n,p_1)\}$ and $Y\sim Binomial \{(m,p_2)\}$. We aim at constructing CIs for $p_1$ (respectively, $\Delta:=p_1-p_2$) for the one- (respectively, two-) sample cases. 
In the one-sample case, we define $C_1$ to be the collection of all confidence intervals, i.e.,
\[
C_1 :=\{[l_{x},u_{x}]\}_{x\in\{0,1,\ldots,n\}}, 
\]
where $l_x, u_x$ is the lower and upper limit of the confidence interval when $X=x$ is observed. Correspondingly, for the two-sample case, we define
\[
C_2 := \{[l_{x,y},u_{x,y}]\}_{x\in\{0,1,\ldots,n\},y\in\{0,1,\ldots,m\}},
\]
and here $[l_{x,y},u_{x,y}]$ is the confidence interval for $\Delta$ when $(X=x,Y=y)$ is observed.

We aim to find an optimal exact CI, where optimality is with respect to the sum of all interval lengths.
In the one-sample case, the length is 
\[
Length(C_1)=\sum_{x=0}^{n} (u_{x}-l_{x}),
\]
and in the two-sample case, it is
\[
Length(C_2)=\sum_{y=0}^{m} \sum_{x=0}^{n} (u_{(x,y)}-l_{(x,y)}).
\]
For computational reasons, we define
a grid  $D$ for $\Delta$ values, and 
a grid $P$ for $p_1, p_2$ single proportions values, 
e.g., $P=\{0,0.01,0.02\ldots ,1\}, 
D=\{-1,-0.99,\ldots 0,0.01,0.02\ldots ,1\}$).
The grids choices are connected to each other since only $(p_1,p_2) \in P \times P$ such that
$p_1 -p_2 \in D$ are active in the problem. {From a statistical point of view, the finer is the grid the better; however, a finer grid comes at the cost of computational burden.} 

The optimization problem we aim to solve for the one-sample case is
\begin{equation}\label{eq:one-opt}
\min_{C_1} Length(C_1) \text{ subject to } P_{p_1} ( p_1 \in [l_X,u_X]) \ge 1- \alpha ~~ \forall p_1 \in P,
\end{equation}~
where the sub-index $p_1$ means that the probability is under $X\sim p_1$ (and similar notation is used for the two-sample case).  For the two-sample case, the optimization problem is
\begin{multline}\label{eq:two-opt}
\min_{C_2} Length(C_2) \\
\text{ subject to } P_{p_1,p_2} (  \Delta \in [l_{(X,Y)},u_{(X,Y)}]) \ge 1- \alpha \text{ for all } (p_1,p_2) \in P \times P \text{ such that } p_1-p_2=\Delta \in D.
\end{multline}

\section{Generalization of the Blyth and Still algorithm to the two-sample case} \label{BSG}

The Blyth and Still algorithm finds all the solutions to the problem \eqref{eq:one-opt}. In Section \ref{sec:BS} the algorithm is described in detail. Generalization of the algorithm to the two-sample case is discussed in Section \ref{sec:BS_gen}. It is shown that the generalized algorithm provides confidence regions rather than intervals.   

\subsection{The Blyth and Still algorithm}\label{sec:BS}

We consider the one-sample case, that is, Problem \eqref{eq:one-opt}, and describe the Blyth and Still algorithm. First, a few definitions are given. 
\begin{definition}\label{def:one-sample}
\begin{itemize}
	\item A subset $S_1=\{r,r+1,\ldots,t\}$ where $0 \le r < t \le n$ is an \textcolor{black}{acceptance region} with respect to $p_1$ if $P_{p_1}(X \in S_1) \geq 1-\alpha$.
	\item A subset $S_1$ is a minimal \textcolor{black}{acceptance region} (henceforth \textcolor{black}{MAR}) with respect to $p_1$, denoted by $\textcolor{black}{MAR}(p_1)$, if there is no other \textcolor{black}{acceptance region} with respect to $p_1$ that has fewer elements. 
	\item 
	Let $S_1 ,\tilde{S}_1$ be two \textcolor{black}{MAR}s with respect to $p_{1}$ and $\tilde{p}_{1}$, where $p_{1}\leq \tilde{p}_{1}$.
	We say that the pair 
	$(S_1,\tilde{S}_1)$ maintains monotonicity if $\min\{S_1\}\leq \min\{\tilde{S}_1\}\text{ and } 
	\max\{S_1\}\leq \max\{\tilde{S}_1\}$. 
\end{itemize}	
\end{definition}

The algorithm can be described as follows:

\begin{tcolorbox}
\begin{center}
	The Blyth and Still algorithm
\end{center}
\noindent Input: $P=\{\rho_1 ,{\rho_2,} \ldots, \rho_{|P|} \}$ - a grid of values in $[0,1]$ such that $\rho_1  \le \rho_2  \le \cdots \le \rho_{|P|}$; $n$ - sample size; $1-\alpha$ - desired level.\\
Output: $C_1$ - a collection of $n+1$ confidence intervals.
\begin{enumerate}
	\item \underline{Find all \textcolor{black}{MAR}s}. For all $p_1 \in P$ calculate all \textcolor{black}{MAR}s. 
	\item \underline{Remove \textcolor{black}{MAR}s that do not maintain monotonicity}. For all $i=1 \tbl{,2},\ldots, |P|-1|$ and for all $S_1=\textcolor{black}{MAR}(\rho_i)$:
	if for all $\tilde{S}_1$ that is a \textcolor{black}{MAR} of $\rho_{i+1}$ the pair $(S_1,\tilde{S}_1)$ does not maintain monotonicity, then remove $S_1$. Also, for all $i=2\tbl{,3} ,\ldots, |P|$ and for all $\tilde{S}_1=\textcolor{black}{MAR}(\rho_i)$, if for all ${S}_1$ that is a \textcolor{black}{MAR} of $\rho_{i-1}$ the pair $(S_1,\tilde{S}_1)$ does not maintain monotonicity, then remove $\tilde{S}_1$.
	\item \underline{Choose linear ordering}. For $i=1$ choose $\textcolor{black}{MAR}^*(\rho_1)$ from all the \textcolor{black}{MAR}s of $\rho_1$ that remained after the previous step. For $i=2,\tbl{3,}\ldots, |P|$,
	choose $\textcolor{black}{MAR}^*(i)$ from all the remaining \textcolor{black}{MAR}s of $\rho_i$ such that  $(\textcolor{black}{MAR}^*(\rho_{i-1}),\textcolor{black}{MAR}^*(\rho_i))$ maintains monotonicity.
	\item \underline{Invert}. For all 
	$x = 0,1, \ldots\tbl{,} n$, define $CR(x):= \{ p_1 \in P : x \in \textcolor{black}{MAR}^*(p_1) \}$ and $l_x:=\min\{ CR(x)\}$, and $u_x:=\max\{CR(x)\}$.
	\item Return $C_1 =\{[l_{x},u_{x}]\}_{x\in\{0,1,\ldots,n\}}$.
\end{enumerate}
\end{tcolorbox}

We now discuss every step of the algorithm in detail.

\noindent 1. \hspace{1mm}\textbf{Find all \textcolor{black}{MAR}s}

\noindent Finding all \textcolor{black}{MAR}s with respect to $p_1$ can be done in the following manner:
set $r=0$ and find the smallest $t_0$ that makes the interval $[0,t_0]$ cover $p_1$ with probability of at least $1-\alpha$, 
i.e., $P_{p_1}( X \in [0,t_0]) \ge 1-\alpha $. 
Then, repeat this procedure for $r=1,\tbl{2,} \ldots, n$: for each $r$, find the smallest integer $t_r$ such that  $P_{p_1}( X \in [r,t_r]) \ge 1-\alpha $.  Notice that there exists a critical value $R$ such that for $r \ge R$ there is no $t_r$ that 
provides coverage of $p_1$ with the desired probability, that is, even if we set $t_r=n$, the interval $S_1=[r,n]$ is not an \textcolor{black}{acceptance region} for $p_1$, i.e.,
$P_{p_1}( X \in [r,n]) < 1-\alpha $.
After calculating $t_0,t_1,...$, the lengths of $[0,t_{0}], [1,t_{1}],...$ are compared and the intervals with minimal length are chosen. Thus, for each $p_1 \in P$ there are $O(n^2)$ calculations, and the total number of calculations in this step is $|P|O(n^2)$.

\noindent 2.\hspace{1mm}\textbf{Remove solutions that do not maintain monotonicity}\\
This step is needed to ensure that $CR(x)$ in the invert step (\# 4) would be an interval rather than a confidence set.  
As mentioned in the introduction, the Sterne CI can lead to optimal confidence sets, which are optimal in terms of length, but they are not necessarily intervals. For a concrete example, suppose that for $p_1=0.1$ the only \textcolor{black}{MAR} is $\textcolor{black}{MAR}(0.1)=[1,7]$ and for $p_1=0.11$ the \textcolor{black}{MAR}s are $\textcolor{black}{MAR}(0.11)=[0,7], [1,8], [2,9]$. Then, the first \textcolor{black}{MAR} $[0,7]$ is removed as it violates the monotonicity assumption with respect to the \textcolor{black}{MAR} $[1,7]$ of $p_1=0.1$. If for $p_1=0.1$ there was more than one \textcolor{black}{MAR}, $[0.7]$ is removed only if it violates the monotonicity assumption for any \textcolor{black}{MAR} of $p_1=0.1$.

\noindent 3.\hspace{1mm}\textbf{Choose linear ordering}\\
There are different ways to choose a linear ordering that will lead to different CIs. However, all of them will be optimal in the sense of the optimization problem in \eqref{eq:one-opt}. 
Blyth and Still explored a few options for choosing \textcolor{black}{MAR}s that have other desired properties. 
For example, if one wants to avoid CIs where $l_x=l_{x+1}$ for some $x$'s, then certain linear orderings should be avoided.

\noindent 4. \hspace{1mm}\textbf{Invert}\\
By the monotonicity property, the set $CR(x)$ is an interval, i.e., there are no holes in $CR(x)$. By the construction of $CR(x)$ we have that $\sum_{x=0}^n \#\{ CR(x) \}=\sum_{p_1 \in P} \#\{ \textcolor{black}{MAR}^*(p_1)\}$, where $\# A$ is the number of elements in set $A$. Since the number of elements in each $ \textcolor{black}{MAR}^*(p_1)$ is minimal, so is $\sum_{x=0}^n \#\{ CR(x) \}$. Minimizing $\sum_{x=0}^n \#\{ CR(x) \}$ is equivalent to Problem \eqref{eq:one-opt} and hence the output of the algorithm is a solution to Problem \eqref{eq:one-opt}. Moreover, by choosing different linear orderings in Step 3, all the optimal solutions can be found by this algorithm.

\subsection{A Generalization of the Blyth and Still algorithm to the two-sample case} \label{sec:BS_gen}

In this section, we consider a generalization of the Blyth and Still algorithm that aims to address Problem \eqref{eq:two-opt}. While the minimal length and the desired coverage probability  are still preserved, we will show that the output of this generalized algorithm is not necessarily a confidence interval, but rather a confidence set. We start with a definition that parallels Definition \ref{def:one-sample}. 

\begin{definition}\label{def:two-sample}
\begin{itemize}
	\item A subset $S_2 \subseteq \{0,1,\ldots,n\} \times \{0,1,\ldots,m\}$  is an \textcolor{black}{acceptance region} with respect to $\Delta \in D$  if for all $(p_1,p_2) \in P \times P$ such that $p_1-p_2 =\Delta$ we have that $P_{p_1,p_2}((X,Y) \in S_2) \geq 1-\alpha$.
	\item A subset $S_2$ is a minimal \textcolor{black}{acceptance region} (henceforth \textcolor{black}{MAR}) with respect to $\Delta \in D$, denoted by $\textcolor{black}{MAR}(\Delta)$, if there is no other \textcolor{black}{acceptance region} with respect to $\Delta$ that has fewer elements.  
\end{itemize}	
\end{definition}

Notice that here we define an \textcolor{black}{acceptance region} to be a subset of $\{0,1,\ldots,n\} \times \{0,1,\ldots,m\}$, without requiring that there are no holes (e.g., $S_2$ in which $(0,2), (0,4) \in S_{2}, (0,3)\notin S_{2}$ is a possible \textcolor{black}{acceptance region}) as in the one sample definition of an \textcolor{black}{acceptance region}.{The motivation is to allow for flexibility in the set of all possible MARs with the hope that a certain choice of MARs will lead to a confidence interval in the inversion step.}
{However, later we demonstrate that there are cases in which all possible choices of} \textcolor{black}{MARs lead to confidence regions that are not intervals.} 

\begin{tcolorbox}
\begin{center}
	The generalized Blyth and Still algorithm
\end{center}
\noindent Input: $P$ - a grid of values in $[0,1]$; $D$ - a grid of values in $[-1,1]$; $n,m$ - sample sizes; $1-\alpha$ - desired level.\\
Output: $\tilde{C}_2$ - a collection of $(n+1)(m+1)$ confidence sets.
\begin{enumerate}
	\item \underline{Find one \textcolor{black}{MAR} for each $\Delta \in D$}. For all $\Delta \in D$ find one \textcolor{black}{MAR}, denoted by $\textcolor{black}{MAR}(\Delta)$. 
	\item \underline{Invert}. For all 
	$(x,y) \in \{0,1,\ldots,n\} \times \{0,1,\ldots,m\}$, define $CR(x,y):= \{ \Delta \in D \text{ if } (x,y) \in \textcolor{black}{MAR}(\Delta) \}$.
	\item Return $\tilde{C_2} =\{CR(x,y)\}_{x\in\{0,1,\ldots,n\},y\in \{0,1,\ldots,m\}}$.
\end{enumerate}
\end{tcolorbox}

Notice that in this algorithm the steps of removing \textcolor{black}{MAR}s that do not maintain monotonicity and choosing linear ordering are not present. This will be explained below, but first, we describe how to find \tbl{the} \textcolor{black}{MAR}s in Step 1. 

Finding \tbl{the} \textcolor{black}{MAR}s in the two-sample case is more complicated than the one-sample equivalent task because one needs to ensure $1-\alpha$ coverage for all $(p_1,p_2) \in P \times P$ that satisfy $p_1-p_2=\Delta$ and not for just one specific $p_1$. Also, in the one-sample case, the \textcolor{black}{MAR}s are intervals but here the \textcolor{black}{MAR}s are general sets.
We found no simple algorithm to compute \tbl{the} \textcolor{black}{MAR}s in the two-sample setting and this step is performed by solving Optimization Problem \ref{OP:1}, which is given below. {This optimization problem consists of $(n+1)(m+1)$ binary variables and has at most $|P|$ constraints for maintaining the coverage probability. In some instances, the solution is not unique. We show below that, unlike the one-sample case, here there is no way \tbl{of choosing a solution that satisfies that $CR(x,y)$} in the invert step is always an interval. Therefore, we selected one solution arbitrarily.}
The optimal solution was computed by a procedure in \tbl{the R software} \citep{R} that uses the Gurobi package \citep{gurobi}. 

\begin{OP} \label{OP:1}
\textbf{Problem parameters}: $D$ - a grid of values in $[-1, 1]$ for $\Delta$;
$P$- a grid of values in $[0,1]$ for $p_{1}$ and $p_{2}$;
$(n,m)$ - number of trials from each sample; confidence coefficient $1-\alpha$.\\
\textbf{Decision variables:} $r(x,y)$ - a binary variable that equals 1 iff $(x,y)$ belongs to the \textcolor{black}{MAR}.\\
\textbf{Objective function:} Minimize $\sum_{y=0}^{m} \sum_{x=0}^{n} r(x,y)$.\\
\textbf{Constraints:}\\
a. Maintain the coverage of $\Delta$\tbl{:}
\begin{equation} \label{eq:sum}
\sum_{y=0}^{m} \sum_{x=0}^{n} r(x,y) \binom{n}{x} \binom{m}{y}
{p_1}^x (1-p_1)^{n-x} {p_2}^y (1-p_2)^{m-y}
\geq 1-\alpha, 
\end{equation}
$\text{ for all }(p_1,p_2) \in P \times P \text{ such that }p_1-p_2=\Delta$.\\
b. The decision variables $r(x,y)$ are binary\tbl{:}
\[
r(x,y) \in \{0,1\} \text{ for all } (x,y) \in \{0,1,\ldots ,n\} \times\{0,1,\ldots ,m\}.
\]
\end{OP}

Furthermore, we used the above program to find all possible solutions for the Optimization Problem \ref{OP:1}. This allows us to show that there are examples in which no ordering of \textcolor{black}{MAR}s will lead to confidence intervals as in the one-dimensional case.   
For example, when $n=5, m=5, \alpha=0.1, P=\{ 0,0.0001,0.0002, \ldots, 1\}$ we find that:\\
a. For $\Delta=-0.4$ the only \textcolor{black}{MAR} contains $(x,y)=(0,5)$.\\
b. For $\Delta=-0.37$ there are five \textcolor{black}{MAR}s, all of them contain $(x,y)=(0,5)$.\\
c. For $\Delta=-0.38$ the only \textcolor{black}{MAR} does not contain $(x,y)=(0,5)$.\\
This means that for any choice of \textcolor{black}{MAR}s in this setting, if $(x,y)=(0,5)$ is observed, the confidence set of the invert step will contain $\Delta=-0.4,-0.37$ but not $\Delta=-0.38$. That is, the optimal confidence set will be composed of at least two disjoint intervals. \textcolor{black}{Furthermore, by examining the constraint \eqref{eq:sum} for continuous $p_1,p_2$ using analytical graphical tools, we observe that this phenomenon still occurs even for a finer grid.} 
The full list of \textcolor{black}{MAR}s for this example are presented in the appendix.

We found that this phenomenon occurs quite often: from six pairs of sample sizes\\ $(n,m) \in \{(10,5),(5,5),(6,4),(9,6),(7,7)\}$
and $\alpha=0.05$, only $(10,5)$ and $(6,4)$ do not have \textcolor{black}{MAR}s with this deficiency. 

It follows that one cannot achieve CIs with minimal length using the Blyth and Still method. Rather, this method guarantees confidence sets (not necessarily intervals) that have a minimal number of elements in $D$ and have the desired coverage level $1-\alpha$.

In section \ref{sec:comparsions} we examine the performance of this method where gaps in the confidence sets are simply filled in order to achieve a confidence interval. 

\section{Performing full optimization} \label{sec:full}

In the previous section we showed that the generalized Blyth and Still algorithm to the two-sample case leads to confidence regions that are optimal in their size, but can be composed of several disjoint intervals, instead of one interval. The solution of filling the gaps between the disjoint intervals is later examined. 

Therefore, a different optimization method should be considered in order to solve Problem \eqref{eq:two-opt}. The aim is to find a set of confidence regions that 
are optimal in length, have the right coverage level, and are constrained to be intervals. This can be done by solving the following optimization problem.
\textcolor{black}{For some instances, the solution is not unique, and when this is the case, we select an arbitrary optimal solution. }

\begin{OP} \label{OP:2}
\textbf{Problem parameters}: $D$ - a grid of values in $[-1, 1]$ for $\Delta$;
$P$- a grid of values in $[0,1]$ for $p_{1}$ and $p_{2}$;
$(n,m)$ - number of trials from each sample; confidence coefficient $1-\alpha$.\\
\textbf{Decision variables:} $l_{(x,y)},u_{(x,y)}$ - the lower and upper limits for when $(x,y)$ is observed;  $r(x,y,\Delta)$ -  a binary variable that equals 1 iff the CI includes $\Delta$ when $(x,y)$ is observed.\\
\textbf{Objective function:}
\text{Minimize} $\sum_{y=0}^{m} \sum_{x=0}^{n}{(u_{(x,y)}-l_{(x,y)})}$.\\
\textbf{Constraints:}\\
a. Maintain the coverage of $\Delta$\tbl{:}
\begin{equation} \label{eq:con1}
\sum_{y=0}^{m} \sum_{x=0}^{n} r(x,y,\Delta) \binom{n}{x} \binom{m}{y}
{p_1}^x (1-p_1)^{n-x} {p_2}^y (1-p_2)^{m-y}
\geq 1-\alpha, 
\end{equation}
$\text{ for all }(p_1,p_2) \in P \times P \text{ such that }p_1-p_2=\Delta$.\\
b. Connecting the variables $r(x,y,\Delta)$ and  $l_{(x,y)}$ and $u_{(x,y)}$: 
\begin{equation}\label{con2}
r{(x,y,\Delta)} \leq \frac{{\Delta-l_{(x,y)}}}{2}+1
\text{ and } r{(x,y,\Delta)} \leq \frac{{u_{(x,y)}-\Delta}}{2}+1
\end{equation}
for all $(x,y,\Delta) \in \tbl{\{0,1,\ldots,n\} \times \{0,1,\ldots,m\}} \times D$.\\
c. Connecting further the variables $r(x,y,\Delta)$ and  $l_{(x,y)}$ and $u_{(x,y)}$:
\begin{equation}\label{con5}
\frac{{u_{(x,y)}-l_{(x,y)}}}{d_{max}}+1 \leq \sum_{ \Delta \in D}{}
r(x,y,\Delta)\leq \frac{{u_{(x,y)}-l_{(x,y)}}}{d_{min}}+1
\end{equation}
for all $(x,y) \in \tbl{\{0,1,\ldots,n\} \times \{0,1,\ldots,m\}}$,
where $d_{min}$ and $d_{max}$ are the minimal and maximal distances between successive  elements in the sorted grid $D$.\\ 
d. The variables $r(x,y,\Delta)$ are binary: 
\[
r{(x,y,\Delta)} \in \{0,1\} \text{ for all }
(x,y,\Delta) \in \tbl{ \{0,1,\ldots,n\} \times \{0,1,\ldots,m\}} \times D.
\]
e. Interval limits are between $[-1,1]$:
\[
-1 \leq l_{(x,y)} \leq 1 \text{ and} 
-1 \leq u_{(x,y)} \leq 1
\text{ for all }
(x,y) \in \tbl{\{0,1,\ldots,n\} \times \{0,1,\ldots,m\}} .
\]
\end{OP} 	
A solution to Optimization Problem \ref{OP:2} finds the shortest CI that has $1-\alpha$ coverage for every $\Delta \in D$, i.e., it solves Problem \eqref{eq:two-opt}. The optimization problem
consists of $2(n+1)(m+1)$ variables that assume values in $D$, and $|D|(n+1)(m+1)$ binary variables.

The constraint in \eqref{con2}  consists of two conditions, which force $r(x,y,\Delta)$ to be 0 if $\Delta< l(x,y)$ or $\Delta> u(x,y)$. 
This is because
\[
\Delta< l(x,y) \Longleftrightarrow  
\frac{{\Delta-l_{(x,y)}}}{2}<0 \Longleftrightarrow  
\frac{{\Delta-l_{(x,y)}}}{2}+1<1.
\]
Thus, by condition \eqref{con2}, if $\Delta< l(x,y)$ then $r(x,y,\Delta)<1$. \textcolor{black}{In addition, the expression $\frac{{(\Delta-l_{(x,y)})}}{2}+1$ is non-negative, and hence $r(x,y,\Delta)<c$ for $c\geq 0$,} which implies that $r(x,y,\Delta)=0$, as it is a binary variable. Similarly, if $\Delta> u(x,y)$, then $r(x,y,\Delta)=0$. If neither $\Delta< l(x,y)$ nor $\Delta> u(x,y)$ is satisfied, then Constraint \eqref{con2} does not restrict $r(x,y,\Delta)$ to a certain value. This is where Constraint \eqref{con5} comes into play. In the case where the grid $D$ is equally-spaced, Constraint \eqref{con5} simplifies to
\begin{equation}\label{con5a}
\sum_{\Delta \in D}{}
r(x,y,\Delta)= \frac{u_{(x,y)}-l_{(x,y)}}{d}+1,   
\end{equation}
where $d$ is the constant difference between successive  elements in the sorted grid $D$. In this case, \eqref{con5a} implies that if $l_{(x,y)} \leq \Delta \leq u_{(x,y)}$, then  $r(x,y,\Delta)=1$. Combining this with \eqref{con2}, we have that the $r$ variables are fully determined by the $l$ and $u$ variables. Constraint \eqref{con5} does not change the optimal value, but rather drastically decreases the number of feasible solutions and thus reduces the number of computations needed to solve Optimization Problem \ref{OP:2}.

Another way of forcing $r{(x,y,\Delta)}$ to be 1 if  $l_{(x,y)} \leq \Delta \leq u_{(x,y)}$, even when $D$ is not equally-spaced, is to 
change the objective function to
\[
\text{minimize } \sum_{y=0}^{m} \sum_{x=0}^{n} (u_{(x,y)}-l_{(x,y)})
-\frac{d_{min}}{2N}\sum_{x=0}^{n}\sum_{y=0}^ {m}\sum_{\Delta \in D}
r(x,y,\Delta),
\]
where $N=(n+1)(m+1)|D|$ is the number of $r$ variables and $d_{min}$ is the minimal distance between consecutive elements in the sorted grid $D$.

If one wishes to find a solution that maintains the symmetry of the binomial distribution under the transformation $p \mapsto 1-p$,
then one can add the restriction
\begin{equation} \label{eq:con6}
u_{(x,y)}=-l_{(n-x,m-y)} \text{ for all } (x,y) \in \tbl{\{0,1,\ldots,n\} \times \{0,1,\ldots,m\}}. 
\end{equation}

In the Generalized Blyth and Still algorithm that was given in Section \ref{sec:BS_gen}, Optimization problem \ref{OP:1} is being solved $|D|$ times, each with 
$(n+1)(m+1)$ binary variables. Here, on the other hand, there are  $|D|(n+1)(m+1)$
binary variables and the optimization problem is solved only once.
Since the running time of the optimization problem solver is not linear in the number of  the binary variables, Optimization Problem \ref{OP:2} is computationally much more difficult.

\section{Comparisons} \label{sec:comparsions}

In this section, we compare the full optimization algorithm of Section \ref{sec:full} and the generalized Blyth and Still algorithm of Section \ref{sec:BS} to several existing methods, both approximate and exact. 

\subsection{A list of methods} \label{sec:list}
The existing methods we have compared are listed below.

\noindent 1. The Wald CI, i.e.,
\[
\hat{p}_{1}-\hat{p}_{2} \pm  z_{1-\frac{\alpha}{2}} \sqrt{\frac{\hat{p}_{1}(1-\hat{p}_{1})}{n}+
\frac{\hat{p}_{2}(1-\hat{p}_{2})}{m}.
}
\]
It is included in our comparison due to its widespread use even though it is known to perform poorly.

\noindent 2. The adjusted Wald CI of  \citet{agresti2000simple} (AC) is given by
\[
\bar{p}_{1}-\bar{p}_{2} \pm  z_{1-\frac{\alpha}{2}} \sqrt{\frac{\bar{p}_{1}(1-\bar{p}_{1})}{n}+
\frac{\bar{p}_{2}(1-\bar{p}_{2})}{m}.
},
\]
where $\bar{p}_{1}=(x+1)/(n+2), \bar{p}_{2}=(y+1)/(m+2)$

\noindent 3. The hybrid score (HS) of \citet{newcombe1998interval}.
\begin{tcolorbox}
\begin{center}
Newcombe hybrid score (HS)
\end{center}
Input: $n,m$ - sample sizes; $1-\alpha$ - confidence coefficient.\\
Output: ${C}_2$ - a collection of $(n+1)(m+1)$ confidence intervals.
\begin{enumerate}
\item \underline{Calculate lower and upper bounds}.
Let $\hat{p}_1=x/n$ and $\hat{p}_2=y/n$. For each $x\in \{0,1,\ldots,n\}$, let 
$l_x(1), u_x(1)$ be the two solutions for $p_1$ of $z_{1-\frac{\alpha}{2}}=\frac{|\hat{p}_1-p_1|}{\sqrt{\frac{p_{1}(1-p_1)}{n}}}$, and for each $y\in \{0,1,\ldots,m\}$, let $l_y(2), u_y(2)$ be the two solutions for $p_2$ of $z_{1-\frac{\alpha}{2}}=\frac{|\hat{p}_2-p_2|}{\sqrt{\frac{p_{2}(1-p_2)}{m}}}$.

\item
\underline{Hybrid score}\tbl{.} For all$(x,y)\in\{0,1,\ldots,n\} \times \{0,1,\ldots,m\}$ define
\begin{align*}
	l(x,y)&=\hat{p}_1-\hat{p}_2-z_{{1-}\frac{\alpha}{2}}\sqrt{\frac{l_x(1)(1-l_x(1))}{n}+\frac{u_y(2)(1-u_y(2))}{m}} ~~~\text{ and }\\
	u(x,y)&=\hat{p}_1-\hat{p}_2+ z_{{1-}\frac{\alpha}{2}}\sqrt{\frac{u_x(1)(1-u_x(1))}{n}+\frac{l_y(2)(1-l_y(2))}{m}}. 
\end{align*}
\item Return $C_2 =\{[l_{(x,y)},u_{(x,y)}]\}_{(x,y)\in\{0,1,\ldots,n\} \times \{0,1,\ldots,m\}}$.
\end{enumerate}
\end{tcolorbox}
The calculations of the HS CI can be found in the R software in the package \tbl{`DescTools'} \citep{DescTools}.

\noindent 4. The exact method of \citet{agresti2001small} (AM)
\begin{tcolorbox}
\begin{center}
The exact method of \citet{agresti2001small} (AM)
\end{center}
\noindent Input: $P$ - a grid of values in $[0,1]$; $D$ - a grid of values in $[-1,1]$; $n,m$ - sample sizes; $1-\alpha$ - confidence coefficient.Output: ${C}_2$ - a collection of $(n+1)(m+1)$ confidence intervals.
\begin{enumerate}
\item \underline{Calculate scores}.
For any triplet $(x,y,\Delta) \in \{0,1,\ldots,n\} \times \{0,1,\ldots,m\} \times D$ define
\[
Z(x,y,\Delta)=\frac{\left(\frac{x}{n}-\frac{y}{m}-\Delta\right)^{2}}{{\frac{\tilde{p}_1(1-\tilde{p}_1)}{n}+\frac{\tilde{p}_2(1-\tilde{p}_2)}{m}}},
\]
where $\tilde{p}_1, \tilde{p}_2$ are the MLE for $p_1, p_2$ under $p_1-p_2=\Delta$, i.e., they maximize the likelihood $	{p_1}^x (1-p_1)^{n-x} {p_2}^y (1-p_2)^{m-y}$, 
under the constraint $p_1-p_2=\Delta$.
\item \underline{Calculate $\lambda$ values}. For any triplet $(x,y,\Delta) \in \{0,1,\ldots,n\} \times \{0,1,\ldots,m\} \times D$ define
\[
\lambda(x,y,\Delta)=\max\left\{
P_{p_1,p_2} \Big(Z(X,Y,\Delta)\geq Z(x,y,\Delta) \Big) : {(p_1,p_2) \in P \times P  \text{ s.t } p_1-p_2=\Delta} \right\}.
\]
\item \underline{Invert} For all 
$(x,y) \in \{0,1,\ldots,n\} \times \{0,1,\ldots,m\}$ define
$CR(x,y):= \{ \Delta \in D \text{ if } \lambda(x,y,\Delta) >\alpha\} $
and $l_{(x,y)}:=\min\{ CR(x,y)\}, u_{(x,y)}:=\max\{CR(x,y)\}$.
\item Return $C_2 =\{[l_{(x,y)},u_{(x,y)}]\}_{(x,y)\in\{0,1,\ldots,n\} \times \{0,1,\ldots,m\}}$.
\end{enumerate}
\end{tcolorbox}

Notice that similar to the generalized Blyth and Still algorithm, $CR(x,y)$ is not necessarily an interval. Therefore, the confidence interval is defined by the minimum and maximum value of $CR(x,y)$. 

We could not find a code in R that implements the AM algorithm, and therefore we wrote our own code. For calculating the MLEs $\tilde{p}_{1},\tilde{p}_{2}$ in Step 1 we used the function `z2stat' in the package \tbl{`PropCIs'}\citep{PropCIs}; an explicit expression for the MSE is given in \citet{miettinen1985comparative}.

We ran the AM algorithm under two modes, which we denote by AM1 and AM2. The first mode is with the grids{$D=\{-1,-0.99,-0.98,\ldots,1\}$ and $P=\{0,0.01,0.02,\ldots,1\}$, and the second mode is 
with the grids $D=\{-1,-0.999,-0.998,\ldots,1\}$ and $P=\{0,0.001,0.002,\ldots,1\}$.}
The reason for considering the coarser grid of the first mode is to attain a better comparison to the full optimization method, in which the finer grid is computationally infeasible. The AM algorithm is sub-optimal but runs much faster than full optimization and therefore can be computed with a finer grid.

\noindent 5. The generalized Blyth and Still algorithm that is given in Section \ref{sec:BS_gen} (BSG).

{We ran the algorithm where the confidence sets filled if they are not intervals.}
As in the AM method, we considered two possible modes, denoted by BSG1 and BSG2. In the first mode we used the grids{$D=\{-1,-0.99,-0.98,\ldots,1\}$ and $P=\{0,0.02,0.04,\ldots,1\}$.}
In the second mode we used the grid $D=\{-1,-0.999,-0.998,\ldots,1\}$ and a different grid $P$ for every $\Delta \in D$, a choice that improves the performance of the algorithm. Namely, for $\Delta \ge 0$ we define
\[
P_{\Delta}=\{\Delta,\ldots,1 \} \text{ with equal  jumps of } \frac{1-\Delta}{100}
\]
and for $\Delta < 0$ we define
\[
P_{\Delta}=\{0,\ldots,1+\Delta\} \text{ with equal jumps of } \frac{1-\Delta}{100}.
\]
The coverage condition of the algorithm in \eqref{eq:sum} is satisfied for any pair
$(p_1,p_1-\Delta)$ where $p_1 \in P_{\Delta}$.

\noindent 6. The full optimization algorithm presented in Section \ref{sec:full} (Full). 

We ran the algorithm of Section \ref{sec:full} with the grid  $D=\{-1,-0.99,-0.98,\ldots,1\},P=\{0,0.01,0.02,\ldots,1\}$ and denote it by Full1. 
Here we only considered the coarse grid since the computational complexity of the optimization problem is much greater. We ran the problem with the symmetric condition 
\eqref{eq:con6} and found that this restriction does not change the length of the CIs in the optimal solution.

The Gurobi software was given a time limit of two minutes. If the time limit is reached, the best solution is reported, as is the gap between this solution to the current lower bound in terms of percentage. The starting point of the algorithm is based on the output of the AM method.

Since the grid is relatively coarse, there are non-negligible amount of differences $p_1-p_2$ for which $1-\alpha$ coverage is not preserved. We examined two ways to overcome this problem, where the updated limits are denoted by $l_{(x,y)}^*$ \tbl{and} $u_{(x,y)}^*$. 
\begin{enumerate}[(a)]
\item Extending the CIs in each direction by adding or reducing $0.01$ (which is the gap size in the grid we used) (Full2), i.e.,
\[
l_{(x,y)}^*= l_{(x,y)}-0.01 \text{ and }    u_{(x,y)}^*= u_{(x,y)}+0.01.
\]
\item Extending the CIs in each direction by adding or reducing $0.01/2$ (Full3), i.e.,
\[
l_{(x,y)}^*= l_{(x,y)}-0.01/2 \text{ and }   u_{(x,y)}^*= u_{(x,y)}+0.01/2.
\]
\end{enumerate}
In these extensions, the new limits are truncated if they exceed the interval $[-1,1]$.

\subsection{Criteria of performance } \label{sec:criteria}

We compare the methods listed in Section \ref{sec:list} according to the following six criteria.

\noindent \underline{AVG length.} The average length in defined by
\[
\frac{\sum_{y=0}^{m} \sum_{x=0}^{n} (u_{(x,y)}-l_{(x,y)})}{(n+1)(m+1)}.
\]

\noindent \underline{\textcolor{black}{\tbl{PCT} of under-coverage.}} Define the coverage probability  function $CP(p_1,p_2):= P_{p_1,p_2} (  \Delta \in [l_{(X,Y)},u_{(X,Y)}])$. The percentage of under-coverage is 
\[
100 \times \int_0^1 \int_0^1 I(CP(p_1,p_2) < 1-\alpha) d{p_1} dp_2.
\]

\noindent \underline{\textcolor{black}{\tbl{PCT} of substantial under-coverage.}} This is defined by
\[
100 \times \int_0^1 \int_0^1 I(CP(p_1,p_2) < 1-\alpha-0.01) d{p_1} dp_2.
\]

\noindent \underline{\tbl{AVG} deviation.}  This is defined by
\[
10,000 \times \int_0^1 \int_0^1 [1-\alpha - CP(p_1,p_2)]I(CP(p_1,p_2) < 1-\alpha) d{p_1} dp_2.
\]

This expression is the loss for an average pair $(p_1,p_2)$ (assuming a uniform distribution), where the loss for each pair is defined by  the difference between the desired level $1-\alpha$ and the actual coverage level $CP(p_1,p_2)$ when $CP(p_1,p_2)$ is below $1-\alpha$ and zero otherwise. 
The factor 10,000 is used since this loss is relatively small in most of the methods we used.

\noindent \underline{Min CL.} The minimum coverage probability is defined by $\min_{ (p_1,p_2)\in[0,1] \times [0,1] } CP(p_1,p_2)$.

\noindent \underline{AVG CL.} The average  coverage probability is $\int_0^1 \int_0^1 CP(p_1,p_2) d{p_1} dp_2$.

For calculating the above criteria (besides AVG length), we sampled 40,000  pairs $(p_1,p_2)$  from a uniform distribution on $[0,1]\times[0,1]$. This defines a grid ${\cal P}$ in $[0,1]\times[0,1]$. Then the above criteria are computed using this grid. For example, the percentage of under-coverage is evaluated by
\[
100 \times \frac{1}{|{\cal P}|} \sum_{ (p_1,p_2)\in {\cal P}} I(CP(p_1,p_2) < 1-\alpha).
\]

\newpage
\subsection{Results}\label{sec:results}

We calculated the resulting CIs of the methods listed in Section \ref{sec:list} for  three cases of $(n,m)$, namely $ (n,m) \in \{(9,6),(14,7),(10,10)\}$.
For each of them, three different confidence coefficients are considered, $\alpha \in \{0.01,0.05,0.1\}$.
For each set of parameters and a CI method, we computed the six criteria of Section \ref{sec:criteria}.

\begin{table}[h!]
\resizebox{\columnwidth}{!}{%
\begin{tabular}{|l|l|l|l|l|l|l|l|l|l|l|l|}
	\hline
	\multicolumn{1}{|l|}{}      & Method            & WALD     & AC      & HS      & AM1    & AM2    & BSG1   & BSG2   & Full1   & Full2  & Full3  \\ \hline \hline
	\multirow{6}{*}{$\alpha=0.01$} & AVG length        & 0.934    & 1.008   & {\bf 0.921}   & 1.017  & 1.026  & 1.009  & 1.028  & 1.004   & 1.024  & 1.014  \\ \cline{2-12} 
	& PCT of under-coverage      & 100.00\% & 27.11\% & 57.73\% & 4.13\% & 0.23\% & 5.11\% & 0.22\% & 6.25\%  & {\bf 0.00\%} & 0.76\% \\ \cline{2-12} 
	& PCT of substantial under-coverage & 100.00\% & 0.79\%  & 19.89\% & 0.01\% & {\bf 0.00\%} & 0.01\% & 0.00\% & 0.01\%  & {\bf 0.00\%} & 0.01\% \\ \cline{2-12} 
	& AVG  deviation    & 810.8    & 7.3     & 58.1    & 0.4    & {\bf 0.0}    & 0.4    & {\bf 0.0}    & 0.6     & {\bf 0.0}    & {\bf 0.0}    \\ \cline{2-12} 
	& Min CL            & 0.029    & 0.959   & 0.9     & 0.925  & 0.987  & 0.925  & 0.988  & 0.925   & 0.99   & 0.969  \\ \cline{2-12} 
	& AVG CL            & 0.909    & 0.993   & 0.987   & 0.994  & 0.995  & 0.994  & 0.995  & 0.994   & 0.994  & 0.994  \\ \hline \hline
	\multirow{6}{*}{$\alpha=0.05$} & AVG length        & {\bf 0.728}    & 0.776   & 0.745   & 0.797  & 0.807  & 0.789  & 0.802  & 0.779   & 0.799  & 0.789  \\ \cline{2-12} 
	& PCT of under-coverage      & 100.00\% & 16.44\% & 49.91\% & 5.69\% & 0.31\% & 6.51\% & 0.38\% & 9.98\%  & {\bf 0.00\%} & 0.92\% \\ \cline{2-12} 
	& PCT of substantial under-coverage & 100.00\% & 3.69\%  & 21.68\% & 0.53\% & {\bf 0.00\%} & 0.70\% & 0.09\% & 1.95\%  & {\bf 0.00\%} & {\bf 0.00\%} \\ \cline{2-12} 
	& AVG  deviation    & 910.2    & 11.6    & 54.5    & 2.3    & 0.1    & 2.8    & 0.2    & 5.4     & {\bf 0.0}    & 0.1    \\ \cline{2-12} 
	& Min CL            & 0.029    & 0.896   & 0.847   & 0.925  & 0.94   & 0.925  & 0.93   & 0.925   & 0.95   & 0.947  \\ \cline{2-12} 
	& AVG CL            & 0.859    & 0.963   & 0.954   & 0.965  & 0.968  & 0.964  & 0.967  & 0.961   & 0.966  & 0.964  \\ \hline \hline
	\multirow{6}{*}{$\alpha=0.1$}  & AVG length        & {\bf 0.616}    & 0.653   & 0.639   & 0.691  & 0.7    & 0.67   & 0.683  & 0.662   & 0.682  & 0.672  \\ \cline{2-12} 
	& PCT of under-coverage      & 99.26\%  & 17.06\% & 43.56\% & 4.45\% & 0.32\% & 6.72\% & 0.80\% & 10.16\% & {\bf 0.00\%} & 0.63\% \\ \cline{2-12} 
	& PCT of substantial under-coverage & 98.46\%  & 8.90\%  & 28.22\% & 0.58\% & 0.06\% & 1.15\% & 0.22\% & 1.94\%  & {\bf 0.00\%} & {\bf 0.00\%} \\ \cline{2-12} 
	& AVG  deviation    & 904.1    & 31.7    & 71.3    & 2.2    & 0.2    & 3.9    & 0.6    & 6.1     & {\bf 0.0}    & 0.1    \\ \cline{2-12} 
	& Min CL            & 0.029    & 0.731   & 0.808   & 0.851  & 0.885  & 0.851  & 0.874  & 0.851   & 0.9    & 0.894  \\ \cline{2-12} 
	& AVG CL            & 0.81     & 0.92    & 0.909   & 0.928  & 0.932  & 0.924  & 0.93   & 0.92    & 0.929  & 0.925  \\ \hline
\end{tabular}%
}	\caption{\footnotesize 
The performance measures of Section \ref{sec:criteria} for the different methods when $\alpha \in \{0.01,0.05,0.1\}$ and $(n,m)=(9,6)$. \tbl{The best method for the criteria AVG length, PCT of under-coverage, PCT of substantial under-coverage and  AVG  deviation is bald-faced.}}\label{tab:n_m_9_6}
\end{table}

\begin{table}[h!]
\resizebox{\columnwidth}{!}{%
\begin{tabular}{|l|l|l|l|l|l|l|l|l|l|l|l|}
	\hline 
	\multicolumn{1}{|l|}{}      & Method            & WALD     & AC      & HS      & AM1    & AM2    & BSG1   & BSG2   & Full1   & Full2  & Full3  \\ \hline \hline
	\multirow{6}{*}{$\alpha=0.01$} & AVG length        & 0.851    & 0.9     & {\bf 0.832}   & 0.892  & 0.903  & 0.888  & 0.905  & 0.88    & 0.9    & 0.89   \\ \cline{2-12} 
	& PCT of under-coverage      & 100.00\% & 33.33\% & 58.65\% & 8.31\% & 0.39\% & 8.81\% & 0.42\% & 14.68\% & {\bf 0.00\%} & 1.67\% \\ \cline{2-12} 
	& PCT of substantial under-coverage & 98.58\%  & 0.95\%  & 15.50\% & 0.02\% & {\bf 0.00\%} & 0.02\% & {\bf 0.00\%} & 0.02\%  & {\bf 0.00\%} & 0.01\% \\ \cline{2-12} 
	& AVG   deviation   & 632.2    & 11.3    & 46.2    & 0.7    & {\bf 0.0}    & 0.8    & {\bf 0.0}    & 1.4     & {\bf 0.0}    & 0.1    \\ \cline{2-12} 
	& Min CL            & 0.038    & 0.956   & 0.895   & 0.894  & 0.988  & 0.894  & 0.986  & 0.894   & 0.99   & 0.958  \\ \cline{2-12} 
	& AVG CL            & 0.927    & 0.992   & 0.987   & 0.993  & 0.993  & 0.993  & 0.994  & 0.992   & 0.993  & 0.993  \\ \hline \hline
	\multirow{6}{*}{$\alpha=0.05$} & AVG length        & {\bf 0.658}    & 0.69    & 0.666   & 0.704  & 0.716  & 0.693  & 0.708  & 0.682   & 0.702  & 0.692  \\ \cline{2-12} 
	& PCT of under-coverage      & 99.60\%  & 18.51\% & 47.75\% & 6.16\% & 0.24\% & 7.14\% & 0.63\% & 14.90\% & {\bf 0.00\%} & 1.47\% \\ \cline{2-12} 
	& PCT of substantial under-coverage & 97.75\%  & 3.34\%  & 13.88\% & 0.40\% & {\bf 0.00\%} & 0.42\% & 0.01\% & 1.38\%  & {\bf 0.00\%} & {\bf 0.00\%} \\ \cline{2-12} 
	& AVG   deviation   & 738.9    & 11.6    & 39.3    & 2.4    & 0.1    & 2.5    & 0.2    & 6.3     & {\bf 0.0}    & 0.1    \\ \cline{2-12} 
	& Min CL            & 0.038    & 0.907   & 0.863   & 0.894  & 0.941  & 0.894  & 0.939  & 0.894   & 0.95   & 0.946  \\ \cline{2-12} 
	& AVG CL            & 0.876    & 0.961   & 0.954   & 0.963  & 0.966  & 0.961  & 0.965  & 0.958   & 0.963  & 0.961  \\ \hline \hline
	\multirow{6}{*}{$\alpha=0.1$}  & AVG length        & {\bf 0.555}    & 0.58    & 0.568   & 0.61   & 0.619  & 0.585  & 0.604  & 0.578   & 0.597  & 0.588  \\ \cline{2-12} 
	& PCT of under-coverage      & 98.96\%  & 17.71\% & 43.05\% & 4.66\% & 0.38\% & 9.70\% & 0.62\% & 14.98\% & {\bf 0.00\%} & 1.16\% \\ \cline{2-12} 
	& PCT of substantial under-coverage & 97.48\%  & 7.05\%  & 23.90\% & 0.52\% & 0.02\% & 1.75\% & 0.08\% & 3.64\%  & {\bf 0.00\%} & {\bf 0.00\%}\\ \cline{2-12} 
	& AVG   deviation   & 742.5    & 21.0    & 50.4    & 2.2    & 0.1    & 5.6    & 0.3    & 11.2    & {\bf 0.0}    & 0.2    \\ \cline{2-12} 
	& Min CL            & 0.038    & 0.802   & 0.824   & 0.875  & 0.889  & 0.864  & 0.885  & 0.842   & 0.9    & 0.891  \\ \cline{2-12} 
	& AVG CL            & 0.826    & 0.918   & 0.908   & 0.926  & 0.93   & 0.917  & 0.927  & 0.913   & 0.924  & 0.919  \\ \hline
\end{tabular}%
}
\caption{\footnotesize 
The performance measures of Section \ref{sec:criteria} for the different methods when $\alpha \in \{0.01,0.05,0.1\}$ and $(n,m)=(14,7)$. \tbl{The best method for the criteria AVG length, PCT of under-coverage, PCT of substantial under-coverage and  AVG  deviation is bald-faced.}}\label{tab:n_m_14_7}
\end{table}

\begin{table}[h!]
\resizebox{\columnwidth}{!}{%
\begin{tabular}{|l|l|l|l|l|l|l|l|l|l|l|l|}
	\hline 
	\multicolumn{1}{|l|}{}      & Method            & WALD     & AC      & HS      & AM1    & AM2    & BSG1    & BSG2   & Full1   & Full2  & Full3  \\ \hline \hline
	\multirow{6}{*}{alpha=0.01} & AVG length        & 0.84     & 0.878   & {\bf 0.823}   & 0.888  & 0.898  & 0.88    & 0.896  & 0.872   & 0.892  & 0.882  \\ \cline{2-12} 
	& PCT of under-coverage      & 100.00\% & 33.22\% & 55.63\% & 7.34\% & 0.43\% & 7.42\%  & 0.61\% & 14.01\% & {\bf 0.00\%} & 1.52\% \\ \cline{2-12} 
	& PCT of substantial under-coverage & 99.26\%  & 0.60\%  & 15.60\% & 0.01\% & {\bf 0.00\%} & 0.01\%  & {\bf 0.00\%} & 0.01\%  & {\bf 0.00\%} & 0.01\% \\ \cline{2-12} 
	& AVG deviation     & 499.7    & 8.2     & 45.8    & 0.7    & {\bf 0.0}    & 0.6     & {\bf 0.0}    & 1.1     & {\bf 0.0}    & 0.1    \\ \cline{2-12} 
	& Min CL            & 0.043    & 0.969   & 0.892   & 0.906  & 0.987  & 0.906   & 0.989  & 0.906   & 0.99   & 0.957  \\ \cline{2-12} 
	& AVG CL            & 0.94     & 0.992   & 0.988   & 0.994  & 0.994  & 0.993   & 0.994  & 0.992   & 0.994  & 0.993  \\ \hline \hline
	\multirow{6}{*}{$\alpha=0.05$} & AVG length        & 0.649    & 0.673   & {\bf 0.654}   & 0.69   & 0.7    & 0.68    & 0.698  & 0.673   & 0.692  & 0.682  \\ \cline{2-12} 
	& PCT of under-coverage      & 100.00\% & 21.56\% & 42.91\% & 9.98\% & 0.76\% & 11.27\% & 1.24\% & 14.73\% & {\bf 0.00\%} & 1.24\% \\ \cline{2-12} 
	& PCT of substantial under-coverage & 99.21\%  & 5.31\%  & 20.74\% & 2.27\% & 0.03\% & 0.31\%  & 0.08\% & 1.27\%  & {\bf 0.00\%} & {\bf 0.00\%} \\ \cline{2-12} 
	& AVG deviation     & 584.2    & 14.9    & 51.0    & 5.6    & 0.3    & 3.8     & 0.4    & 5.5     & {\bf 0.0}    & 0.1    \\ \cline{2-12} 
	& Min CL            & 0.043    & 0.907   & 0.835   & 0.906  & 0.936  & 0.906   & 0.939  & 0.906   & 0.95   & 0.946  \\ \cline{2-12} 
	& AVG CL            & 0.892    & 0.96    & 0.954   & 0.962  & 0.965  & 0.961   & 0.965  & 0.959   & 0.964  & 0.961  \\ \hline \hline
	\multirow{6}{*}{$\alpha=0.1$}  & AVG length        & {\bf 0.547}    & 0.566   & 0.556   & 0.588  & 0.598  & 0.575   & 0.594  & 0.566   & 0.586  & 0.576  \\ \cline{2-12} 
	& PCT of under-coverage      & 98.15\%  & 23.24\% & 44.90\% & 9.31\% & 0.74\% & 14.35\% & 1.53\% & 18.32\% & {\bf 0.00\%} & 1.18\% \\ \cline{2-12} 
	& PCT of substantial under-coverage & 92.87\%  & 11.73\% & 26.32\% & 2.72\% & 0.13\% & 4.84\%  & 0.47\% & 7.94\%  & {\bf 0.00\%} & 0.01\% \\ \cline{2-12} 
	& AVG deviation     & 588.1    & 32.5    & 62.6    & 6.3    & 0.4    & 11.9    & 1.3    & 18.5    & {\bf 0.0}    & 0.2    \\ \cline{2-12} 
	& Min CL            & 0.043    & 0.743   & 0.801   & 0.835  & 0.883  & 0.835   & 0.872  & 0.835   & 0.9    & 0.865  \\ \cline{2-12} 
	& AVG CL            & 0.841    & 0.916   & 0.907   & 0.919  & 0.924  & 0.917   & 0.927  & 0.912   & 0.923  & 0.918  \\ \hline
\end{tabular}%
}
\caption{\footnotesize 
The performance measures of Section \ref{sec:criteria} for the different methods when $\alpha \in \{0.01,0.05,0.1\}$ and $(n,m)=(10,10)$. \tbl{The best method for the criteria AVG length, PCT of under-coverage, PCT of substantial under-coverage and  AVG  deviation is bald-faced.}}\label{tab:n_m_10_10}
\end{table}

The results for $(n,m)=(9,6), (14,7), (10,10)$ are given in Tables \ref{tab:n_m_9_6}, \ref{tab:n_m_14_7}, \ref{tab:n_m_10_10}, respectively. A few observations and conclusions are now given.

\begin{itemize}
\item
The WALD CI performs poorly.  Almost for all pairs, the  coverage probability is below the desired level $1-\alpha$ and even below $1-\alpha-0.01$. Also, the average coverage is well below the desired level. This finding is not surprising as the WALD CI relies on asymptotic approximation, which is not valid for small sample sizes.

\item We considered three non-exact methods: WALD, HS and AC. Comparing these methods in terms of average length, the order is \tbl{usually} 
$\text{WALD} < \text{HS} < \text{AC}$, but the same order holds under the
\textcolor{black}{under-coverage and substantial under-coverage criteria}. This means that narrower CIs come with the price of under-coverage.

\item
The Full1 method produces CI with optimal length, or close to optimal; see the discussion below. As we expected, it has the shortest average length among all exact CIs. 
Compared to the approximate CIs it is longer by {$2\% \textendash 10\%$} from HS and WALD and it has a similar length as AC.

\item 
The Full1 method does not guarantee exact coverage for any $(p_1,p_2)$, just for the pairs in the grid. 
For $(n,m)=(9,6)$, the percentage of under-coverage pairs range from $6\%$ \tbl{for} $\alpha=0.01$, to $10\%$ for $\alpha=0.1$.
For the two other sample sizes, it ranges from $14\%$ to $18\%$.
Examining Full1 by the criterion of percentage of substantial under-coverage, we can see that it has good performance, especially for small $\alpha$.
Yet, for $(n,m,\alpha)=(10,10,0.1)$ 
the percentage of substantial under-coverage reaches $8\%$, which might be too high. 
Still, the Full1 has a smaller percentage of under-coverage and percentage of substantial under-coverage compared to the approximate CIs, including AC.

\item
The exact methods Full1, BSG1 and AM1 ran with the same grid for $\Delta$. Among these methods, the order of the average length is \tbl{usually} $\text{Full1}<\text{BSG1}<\text{AM1}$. The length improvement of Full1 compared to AM1 is about{$2\%\textendash5\%$}. On the other hand, AM1 has better coverage than BSG1 and Full1.

\item
The modification of Full2 produces \tbl{a} CI that is exact for all pairs but it comes at the cost of a larger length of 0.02.
Full3 does not guarantee exact coverage, but the percentage of under-coverage is decreased by about $90\%$ compared to Full1, and the intervals are extended by half the amount compared to Full2.

\item
The BSG2 method achieves significant improvement in the coverage criteria compared to BSG1, at the cost of average length that is greater by about $2\%$.  Similarly, AM2 improves AM1 in terms of coverage, but the average length increases slightly.

\item 
\tbl{Out of} all the exact methods examined, only 
BSG2, AM2, Full2 and Full3 have satisfactorily performance for the coverage criteria. The  coverage probability of Full2 is always larger than $1-\alpha$ in the above parameters.
Comparing BSG2, AM2 and Full3
we can observe that Full3 has the largest percentage of under-coverage, for most of the  nine combinations of $n,m$ \tbl{and} $\alpha$ we considered, while AM2 has the lowest. On the other hand, Full3 has the smallest percentage of substantial under-coverage, smaller than $0.01\%$ for all 9 cases. AM2 and BSG2 have slightly higher numbers, yet still very low, ranging from $0.13\%$ to $0.47\%$.

Considering the criterion of AVG deviation, all three methods have low scores, in comparison to the other methods. BSG2 has a slightly higher score than AM2 and Full3, which are
mostly comparable.

\item 
{In addition we checked if the Full3 CIs maintain the nestedness property. We checked for every sample sizes $3\leq m \leq n \leq 15$ and for each sample results $(x,y) \in \{0,1,\ldots,n\} \times \{0,1,\ldots,m\}$ that the $99\%$ CI contains the $95\%$ CI, and the latter contains the $90\%$ CI.
Overall there are {$50$} violations of nestedness out of {$9,191$ comparisons $(0.544\%)$}. For example, when the sample sizes are $(n,m)=(10,7)$ and the  observations are $(x,y)=(8,1)$, the $90\%$ CI is $[0.185,0.855]$, while the $95\%$ CI is $[0.195,0.895]$, which does not contain the former interval.
} 

\end{itemize}

\begin{figure}[H]
\begin{center}
\includegraphics[scale=0.45]{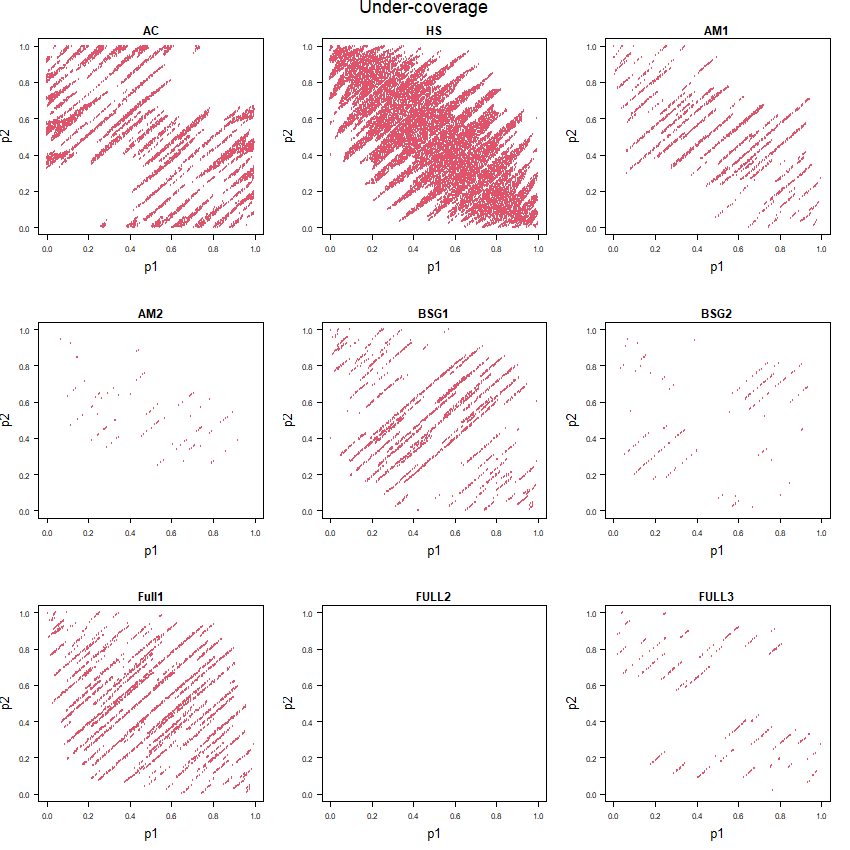}
\caption{\footnotesize Plotting all pairs $(p_1,p_2) \in {\cal P}$ for which $CP(p_1,p_2)$ is below $1-\alpha$ when $(n,m,\alpha)=(9,6,0.05)$ for all methods listed in Section \ref{sec:list} besides WALD. }\label{fig:under}
\end{center}   
\end{figure}

To examine further the under-coverage of the different methods we plotted in Figure \ref{fig:under} all pairs $(p_1,p_2) \in {\cal P}$ for which $CP(p_1,p_2)$ is below $1-\alpha$ when  $(n,m,\alpha)=(9,6,0.05)$ for all methods excluding WALD.

We observe that 
AM2, BSG2 and Full3 have similar low under-coverage, but the pattern is a bit different. For AM2 the under-coverage is mostly for pairs $(p_1,p_2)$ that are close $(\frac{1}{2},\frac{1}{2})$,
while for Full3 it is mostly for large $\Delta=|p_1-p_2|$.
The graph of Full2 is empty as there is no under-coverage for this method.

\begin{figure}[H]
\begin{center}
\includegraphics[scale=0.35]{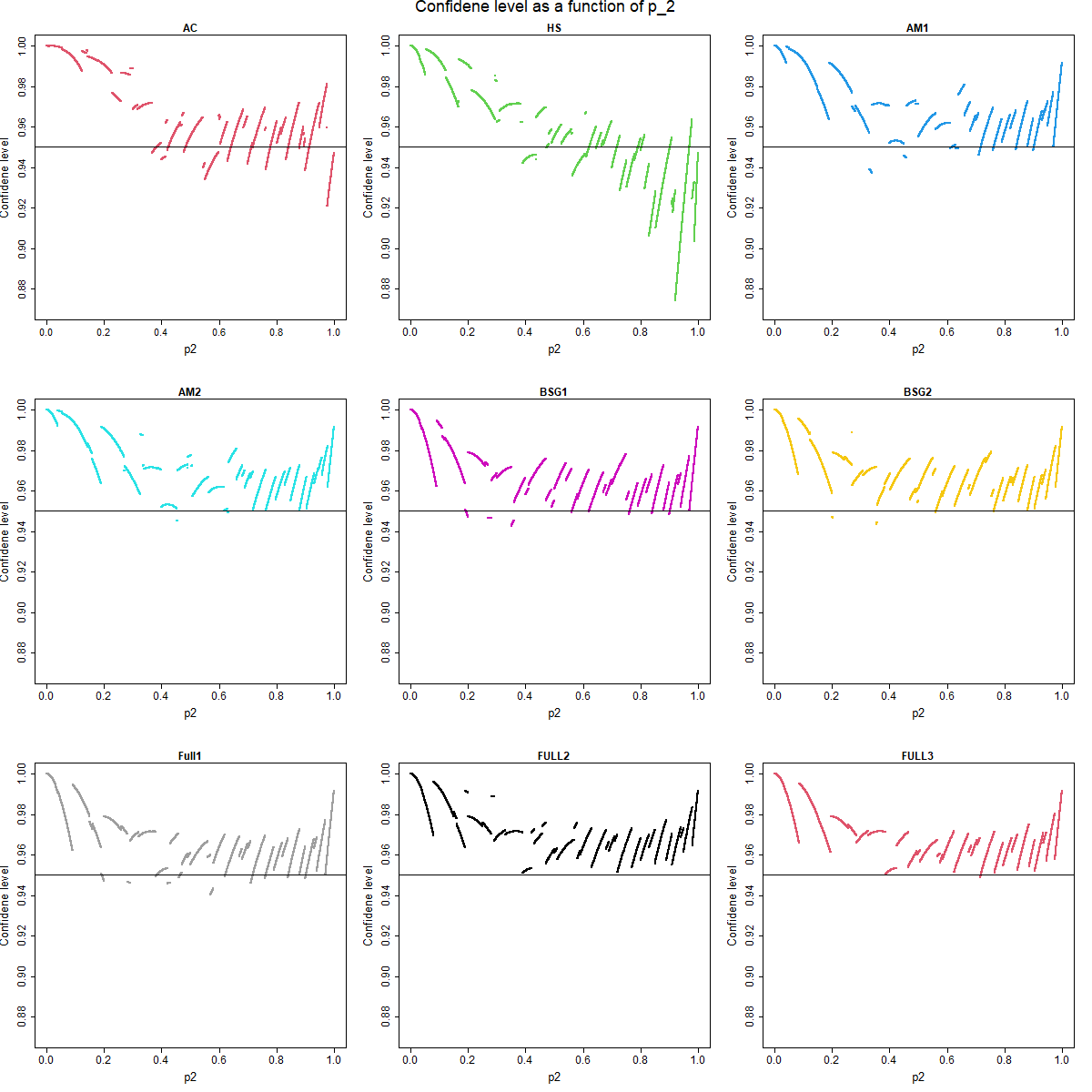}
\caption{\footnotesize Plotting the  coverage probability as a function of $p_2$
	when $p_1=0.5$ and $(n,m,\alpha)=(9,6,0.05)$ for for all methods listed in Section \ref{sec:list} excluding WALD. The vertical line represents the $1-\alpha=0.95$ confidence coefficient.  }\label{fig:p_2}
	\end{center}   
\end{figure}

Additionally, Figure \ref{fig:p_2} plots the  coverage probability as a function of $p_2$ when $p_1=0.5$. We can see that all the seven exact methods (AM1, AM2, BSG1, BSG2, Full1, Full2, Full3)  exhibit a similar pattern, and the  coverage probability is above $1-\alpha$ for almost all $p_2$. Notice that the graph of Full2 has a few short lines (looking like points) in the high-confidence area, which do not exist in the Full3 graph. This is due to the extension of the limits by $0.01/2$ in Full3 compared to the extension of $0.01$ in Full2.

\vspace{2mm}

Table \ref{tab:gap} reports the decrease in the average length of
the solutions found by the Full1 method compared to the best lower bound that was computed. It is demonstrated that the gap between the best lower bound and the solution that was found is quite small. 
Even if the time limit of the algorithm is extended, we believe that it generally would not result in better performance. By observing the outputs of the optimization algorithm throughout the run, it seems that the solution found is optimal or very close to optimal, and more running time will mostly improve the computation of the lower bound, and not the solution itself. 
For example, for
$(n,m,\alpha)=(10,10,0.05)$ the solution after 180 seconds, was the same one that was found  after 30 seconds. The changes were only in the computation of the gap: from $1.67\%$ to $0.87\%$.

\begin{table}[h!]
\begin{center}
\begin{tabular}{lllll}
	&                                &                                 &                                  &  \\ \cline{1-4}
	\multicolumn{1}{|l|}{$\alpha$\textbackslash{}($n,m$)} & \multicolumn{1}{l|}{$(n=9,m=6)$} & \multicolumn{1}{l|}{$(n=14,m=7)$} & \multicolumn{1}{l|}{$(n=10,m=10)$} &  \\ \cline{1-4}
	\multicolumn{1}{|l|}{0.01}                       & \multicolumn{1}{l|}{0.1\%}   & \multicolumn{1}{l|}{0.957\%}    & \multicolumn{1}{l|}{0.922\%}     &  \\ \cline{1-4}
	\multicolumn{1}{|l|}{0.05}                       & \multicolumn{1}{l|}{0.33\%}   & \multicolumn{1}{l|}{0.94\%}    & \multicolumn{1}{l|}{1.08\%}      &  \\ \cline{1-4}
	\multicolumn{1}{|l|}{0.1}                        & \multicolumn{1}{l|}{0.77\%}   & \multicolumn{1}{l|}{1.31\%}    & \multicolumn{1}{l|}{1.33\%}      &  \\ \cline{1-4}
\end{tabular}
\caption{\footnotesize 
	A bound of the gap, in terms of percentage of length, between the optimal solution and the one found by Full1 as computed by the Gurobi package.}\label{tab:gap}
	\end{center}
\end{table}


Considering both coverage and length, it seems that Full3 is the best method among the ones we suggested, namely, Full1, Full2, Full3, BSG1 and BSG2. Among the other methods, AM2 has the best performance. Comparing Full3 and AM2, they perform similarly in the coverage criteria but Full3 has a smaller average length.

\begin{figure}[H]
\begin{center}
\includegraphics[scale=0.85]{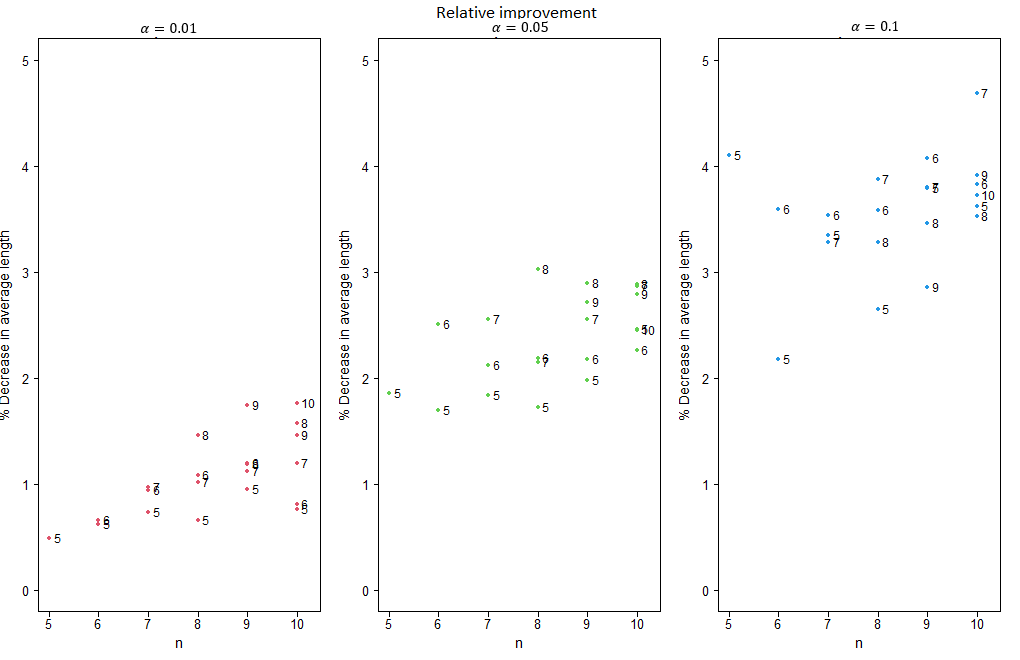}
\caption{\footnotesize Plotting the relative improvement as defined in \eqref{eq:rel} for every pair $(n,m)$ where $n \in \{5,\ldots,10\}$ and $m \in \{5,\ldots,n\}$ and for $\alpha \in \{0.01,0.05,0.1\}$.
	The $x$-axis in the graphs is $n$ and the number near each point is the corresponding $m$.}\label{fig:relative}
	\end{center}   
\end{figure}

To examine further the decrease of length of Full3 compared to AM2, we considered 
21 pairs of $(n,m)$, where  $5 \le m \le n \le 10$.
For each such pair and for $\alpha \in \{0.01, 0.05, 0.1\}$  we computed the relative improvement, which is defined by 
\begin{equation}\label{eq:rel}
100 \times \frac{\text{AVG length(AM2)-AVG length(Full3)}}{\text{AVG length (AM2)}}.    
\end{equation}
The results are plotted in Figure \ref{fig:relative}.  We observe that for all 21 pairs Full3 produced shorter intervals and the relative improvement varies from 0.5\% to 5\%. The larger the $\alpha$, the larger is the relative improvement.  For $\alpha=0.01$, the relative improvement is about  $1\%$, and for $\alpha=0.05$, the range is about 2.5\%  and 4\%, respectively.
It also seems that the relative improvement tends to increase with $n$. In all runs of Full3, the gap between the solution obtained to the lower bound is rather small and the largest gap is $1.35\%$.
Figure \ref{fig:relative_2}, which appears in the appendix, extends Figure \ref{fig:relative} to sample sizes $(n,m)$ where $3 \le m \le n \le 15$.

\textcolor{black}{Figures \ref{fig:relative} and  \ref{fig:relative_2} demonstrate that larger sample sizes lead to more relative improvement. This is because the degrees of freedom of the optimization method are larger for larger sample sizes.  Therefore, the advantage of performing full optimization is more significant.}


\subsection{Summary of the findings}
The Full algorithm was shown to be computationally
feasible for small $n,m$ using the rather coarse grid of $D=\{-1,-0.99,\ldots,1\}$.
While the resulting CIs do not have the right coverage probability for $p_1$ and $p_2$ that are not in the grid, simple adjustments can be made to improve the coverage at a small cost in the average length. The adjusted method, Full3, is comparable, in terms of coverage, to AM2 and BSG2, which are computed under a finer grid, but has shorter CIs.

\section{Discussion} \label{sec:discussion}

For small $n,m$ ($n,m \leq 15)$ we recommended the use of the Full3 method, as it has good coverage and a small  average length.
Tables for various $(n,m,\alpha)$ \textcolor{black}{of the Full3 method} are presented in the following 
\href{https://technionmail-my.sharepoint.com/:f:/g/personal/ap_campus_technion_ac_il/El-213Kms51BhQxR8MmQJCYBDfIsvtrK9mQIey1sZnZWIQ?e=hxGunl}{link}. 
The second best method is the AM2 method, and it can be used when Full3 is not available.

We also tried several examples with larger sample sizes than 15. When both sample sizes were 25, the algorithm could not find feasible solutions. For smaller sample sizes (around 20) the results were similar to what was reported in Section 5.3. However, a more thorough study is required for larger sample sizes, and we leave this for future research.

Extensions of this work can go in several directions. One can consider extending the Full algorithm to other frequently used discrete distributions, like Poisson or Hyper-geometric. This amounts to changing the coverage criterion \eqref{eq:con1}
according to the distribution used. One can also consider other related optimization problems, for example finding the shortest CIs that have an average confidence coefficient of $1-\alpha$, and the minimal  coverage probability is above $1-\beta$ for some $\beta>\alpha$. The availability of powerful optimization algorithms and software allows one to investigate such problems.

\bibliographystyle{chicago}
\bibliography{sample}

\begin{thebibliography}{}

\bibitem[\protect\citeauthoryear{Agresti and Caffo}{Agresti and
  Caffo}{2000}]{agresti2000simple}
Agresti, A. and B.~Caffo (2000).
\newblock Simple and effective confidence intervals for proportions and
  differences of proportions result from adding two successes and two failures.
\newblock {\em The American Statistician\/}~{\em 54\/}(4), 280--288.

\bibitem[\protect\citeauthoryear{Agresti and Coull}{Agresti and
  Coull}{1998}]{agresti1998approximate}
Agresti, A. and B.~A. Coull (1998).
\newblock Approximate is better than “exact” for interval estimation of
  binomial proportions.
\newblock {\em The American Statistician\/}~{\em 52\/}(2), 119--126.

\bibitem[\protect\citeauthoryear{Agresti and Min}{Agresti and
  Min}{2001}]{agresti2001small}
Agresti, A. and Y.~Min (2001).
\newblock On small-sample confidence intervals for parameters in discrete
  distributions.
\newblock {\em Biometrics\/}~{\em 57\/}(3), 963--971.

\bibitem[\protect\citeauthoryear{Blaker}{Blaker}{2000}]{blaker2000confidence}
Blaker, H. (2000).
\newblock Confidence curves and improved exact confidence intervals for
  discrete distributions.
\newblock {\em Canadian Journal of Statistics\/}~{\em 28\/}(4), 783--798.

\bibitem[\protect\citeauthoryear{Blyth and Still}{Blyth and
  Still}{1983}]{blyth1983binomial}
Blyth, C.~R. and H.~A. Still (1983).
\newblock Binomial confidence intervals.
\newblock {\em Journal of the American Statistical Association\/}~{\em
  78\/}(381), 108--116.

\bibitem[\protect\citeauthoryear{Brown, Cai, and DasGupta}{Brown
  et~al.}{2001}]{brown2001interval}
Brown, L.~D., T.~T. Cai, and A.~DasGupta (2001).
\newblock Interval estimation for a binomial proportion.
\newblock {\em Statistical science\/}~{\em 16\/}(2), 101--133.

\bibitem[\protect\citeauthoryear{Brumback and Berg}{Brumback and
  Berg}{2008}]{Brumback2008}
Brumback, B. and A.~Berg (2008).
\newblock On effect-measure modification: relationships among changes in the
  relative risk, odds ratio, and risk difference.
\newblock {\em Statistics in Medicine\/}~{\em 27\/}(18), 3453--3465.

\bibitem[\protect\citeauthoryear{Chan and Zhang}{Chan and
  Zhang}{1999}]{chan1999test}
Chan, I.~S. and Z.~Zhang (1999).
\newblock Test-based exact confidence intervals for the difference of two
  binomial proportions.
\newblock {\em Biometrics\/}~{\em 55\/}(4), 1202--1209.

\bibitem[\protect\citeauthoryear{Clopper and Pearson}{Clopper and
  Pearson}{1934}]{clopper1934use}
Clopper, C.~J. and E.~S. Pearson (1934).
\newblock The use of confidence or fiducial limits illustrated in the case of
  the binomial.
\newblock {\em Biometrika\/}~{\em 26\/}(4), 404--413.

\bibitem[\protect\citeauthoryear{Crow}{Crow}{1956}]{crow1956confidence}
Crow, E.~L. (1956).
\newblock Confidence intervals for a proportion.
\newblock {\em Biometrika\/}~{\em 43\/}(3/4), 423--435.

\bibitem[\protect\citeauthoryear{Fagerland, Lydersen, and Laake}{Fagerland
  et~al.}{2015}]{fagerland2015recommended}
Fagerland, M.~W., S.~Lydersen, and P.~Laake (2015).
\newblock Recommended confidence intervals for two independent binomial
  proportions.
\newblock {\em Statistical methods in medical research\/}~{\em 24\/}(2),
  224--254.

\bibitem[\protect\citeauthoryear{Fay and Hunsberger}{Fay and
  Hunsberger}{2021}]{Fay2021}
Fay, M.~P. and S.~A. Hunsberger (2021).
\newblock {Practical valid inferences for the two-sample binomial problem}.
\newblock {\em Statistics Surveys\/}~{\em 15\/}(none), 72 -- 110.

\bibitem[\protect\citeauthoryear{{Gurobi Optimization, LLC}}{{Gurobi
  Optimization, LLC}}{2023}]{gurobi}
{Gurobi Optimization, LLC} (2023).
\newblock {Gurobi Optimizer Reference Manual}.

\bibitem[\protect\citeauthoryear{Keer}{Keer}{2023}]{Keer2023}
Keer, P. (2023).
\newblock Hypothesis testing in contingency tables: A discussion, and exact
  unconditional tests for r×c tables.
\newblock Available at
  \url{https://repository.tudelft.nl/islandora/object/uuid:7102c72a-ab3e-49a6-9164-131de660053e?collection=education}.

\bibitem[\protect\citeauthoryear{Mart{\'\i}n~Andr{\'e}s, Gay{\'a}~Moreno,
  {\'A}lvarez~Hern{\'a}ndez, and Herranz~Tejedor}{Mart{\'\i}n~Andr{\'e}s
  et~al.}{2024}]{Antonio2024}
Mart{\'\i}n~Andr{\'e}s, A., F.~Gay{\'a}~Moreno, M.~{\'A}lvarez~Hern{\'a}ndez,
  and I.~Herranz~Tejedor (2024).
\newblock Miettinen and nurminen score statistics revisited.
\newblock {\em Journal of Biopharmaceutical Statistics\/}, 1--14.

\bibitem[\protect\citeauthoryear{Miettinen and Nurminen}{Miettinen and
  Nurminen}{1985}]{miettinen1985comparative}
Miettinen, O. and M.~Nurminen (1985).
\newblock Comparative analysis of two rates.
\newblock {\em Statistics in medicine\/}~{\em 4\/}(2), 213--226.

\bibitem[\protect\citeauthoryear{Newcombe}{Newcombe}{1998}]{newcombe1998interval}
Newcombe, R.~G. (1998).
\newblock Interval estimation for the difference between independent
  proportions: comparison of eleven methods.
\newblock {\em Statistics in medicine\/}~{\em 17\/}(8), 873--890.

\bibitem[\protect\citeauthoryear{{R Core Team}}{{R Core Team}}{2021}]{R}
{R Core Team} (2021).
\newblock {\em R: A Language and Environment for Statistical Computing}.
\newblock Vienna, Austria: R Foundation for Statistical Computing.

\bibitem[\protect\citeauthoryear{Santner and Snell}{Santner and
  Snell}{1980}]{santner1980small}
Santner, T.~J. and M.~K. Snell (1980).
\newblock Small-sample confidence intervals for p 1--p 2 and p 1/p 2 in
  2$\times$ 2 contingency tables.
\newblock {\em Journal of the American Statistical Association\/}~{\em
  75\/}(370), 386--394.

\bibitem[\protect\citeauthoryear{Scherer}{Scherer}{2022}]{PropCIs}
Scherer, R. (2022).
\newblock {\em PropCIs: Various Confidence Interval Methods for Proportions}.

\bibitem[\protect\citeauthoryear{Signorell}{Signorell}{2024}]{DescTools}
Signorell, A. (2024).
\newblock {\em DescTools: Tools for Descriptive Statistics}.
\newblock R package version 0.99.55.

\bibitem[\protect\citeauthoryear{Sterne}{Sterne}{1954}]{sterne1954some}
Sterne, T.~E. (1954).
\newblock Some remarks on confidence or fiducial limits.
\newblock {\em Biometrika\/}~{\em 41\/}(1/2), 275--278.

\bibitem[\protect\citeauthoryear{Wilson}{Wilson}{1927}]{wilson1927probable}
Wilson, E.~B. (1927).
\newblock Probable inference, the law of succession, and statistical inference.
\newblock {\em Journal of the American Statistical Association\/}~{\em
  22\/}(158), 209--212.

\end{thebibliography}

\section{Appendix}

\section*{A list of notation}

\begin{itemize}
\item 
$p_1, p_2$ - proportions of binomial distribution

\item
$\Delta=p_1-p_2$ -difference between two proportions   

\item
$1-\alpha$ - the stated confidence coefficient

\item
$n, m$ - the sample sizes 

\item
$z_{1-\frac{\alpha}{2}}$ - the $1-\frac{\alpha}{2}$ quantile of standard normal distribution  

\item
$X$ - random variable $\sim binomial(n,p_1)$,
$Y$ - random variable $\sim binomial(m,p_2)$
\item
$x,y$ - The sample results

\item
$l_x , u_x$ -  lower and upper limit of the confidence interval for $p_1$  when X = x is observed. $l_{(x,y)} , u_{(x,y)}$ lower and upper limit of the confidence interval for  $\Delta$  when $(X,Y) = (x,y)$ is observed

\item
CI- \textcolor{black}{the} confidence interval 

\item
$C_1$- the collection of all confidence intervals for one sample case:
\[
C_1 :=\{[l_{x},u_{x}]\}_{x\in\{0,1,\ldots,n\}} 
\]

\item
$C_2$- the collection of all confidence intervals for two sample case:
\[
C_2 := \{[l_{x,y},u_{x,y}]\}_{x\in\{0,1,\ldots,n\},y\in\{0,1,\ldots,m\}}
\]

\item
$D$- grid for $\Delta$ values

\item
$P$- grid for $p_1, p_2$ values. 
\end{itemize}

\section*{Relative improvement to additional pairs of sample sizes}
\begin{figure}[H]
\begin{center}
	\includegraphics[scale=0.85]{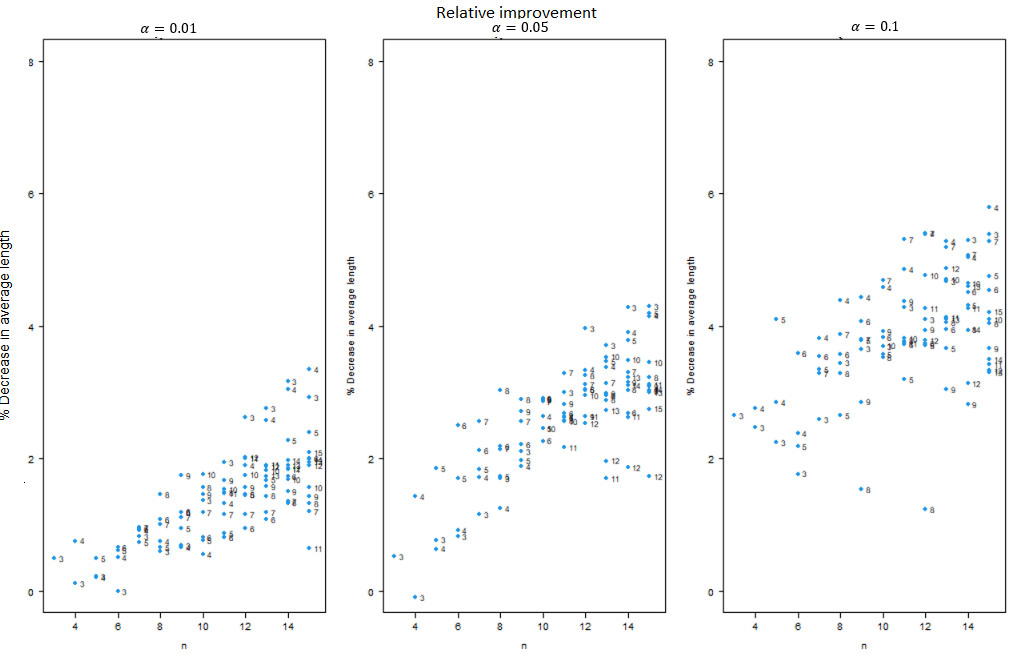}
	\caption{\footnotesize Plotting the relative improvement as defined in \eqref{eq:rel} for every pair $(n,m)$ where $n \in \{3,\ldots,15\}$ and $m \in \{3,\ldots,n\}$ and for $\alpha \in \{0.01,0.05,0.1\}$.
		The $x$-axis in the graphs is $n$ and the number near each point is the corresponding $m$.}\label{fig:relative_2}
\end{center}   
\end{figure}

Figure \ref{fig:relative_2} displays the relative improvement as
shown in figure \ref{fig:relative} to sample sizes $(n,m)$ where $3 \le m \le n \le 15$.

\section*{The full list of {MAR}s for the example presented in section
	\ref{sec:BS_gen}}

\textcolor{black}{MAR} for $\Delta=-0.37$:

$\textcolor{black}{MAR}1(-0.37)=\{(0,1),(0,2),(0,3),(0,4),(0,5),
(1,1),(1,2),(1,3),(1,4),(1,5),
\newline
(2,2),(2,3),(2,4),(2,5),
(3,3),(3,4),(3,5),
(4,4),(4,5)
\}$

\textcolor{black}{MAR}s for $\Delta=-0.4$:

$\textcolor{black}{MAR}1(-0.4)=\{(0,1),(0,2),(0,3),(0,4),(0,5),
(1,1),(1,2),(1,3),(1,4),(1,5),
\newline
(2,2),(2,3),(2,4),(2,5),
(3,3),(3,4),(3,5),
(4,4),(4,5)
\}$

$\textcolor{black}{MAR}2(-0.4)=\{(0,1),(0,2),(0,3),(0,4),(0,5),
(1,1),(1,2),(1,3),(1,4),(1,5),
\newline
(2,3),(2,4),(2,5),
(3,2),(3,3),(3,4),(3,5),
(4,4),(4,5)
\}$

$\textcolor{black}{MAR}3(-0.4)=\{(0,1),(0,2),(0,3),(0,4),(0,5),
(1,1),(1,2),(1,3),(1,4),(1,5),
\newline
(2,1),(2,3),(2,4),(2,5),
(3,3),(3,4),(3,5),
(4,4),(4,5)
\}$

$\textcolor{black}{MAR}4(-0.4)=\{(0,1),(0,2),(0,3),(0,4),(0,5),
(1,1),(1,2),(1,3),(1,4),(1,5),
\newline
(2,2),(2,3),(2,4),(2,5),
(3,2),(3,4),(3,5),
(4,4),(4,5)
\}$

$\textcolor{black}{MAR}5(-0.4)=\{(0,1),(0,2),(0,3),(0,4),(0,5),
(1,1),(1,2),(1,3),(1,4),(1,5),
\newline
(2,2),(2,3),(2,4),(2,5),
(3,4),(3,5),
(4,3),(4,4),(4,5)
\}$

\textcolor{black}{MAR} for $\Delta=-0.38$ :

$\textcolor{black}{MAR}1(-0.38)=\{(0,1),(0,2),(0,3),(0,4),
(1,1),(1,2),(1,3),(1,4),(1,5),
\newline
(2,2),(2,3),(2,4),(2,5),
(3,3),(3,4),(3,5),
(4,4),(4,5)
\}$

\end{document}